\documentclass{iaus238}
\usepackage{graphicx}

\title[Microquasars: disk--jet coupling...] 
{Microquasars: disk--jet coupling in stellar-mass black holes}

\author[I.~F. Mirabel]   
{I. F\'elix Mirabel\,\thanks{On leave from CEA-Saclay, France.\hfill~}}

\affiliation{European Southern Observatory, Alonso de Cordova 3107, Santiago, Chile\\[\affilskip] 
email: fmirabel@eso.org}

\pubyear{2006}
\volume{238}  
\date{to be defined}
\jname{Black Holes: from Stars to Galaxies -- across the Range of Masses}
\editors{V. Karas \& G. Matt}
\begin{document}

\maketitle

\begin{abstract}
Microquasars provide new insights into:  1)~the physics of relativistic
jets from black holes, 2)~the connection between accretion and ejection, 
and 3)~the physical mechanisms in the formation of stellar-mass black
holes. Furthermore, the studies  of microquasars in our Galaxy can
provide in the future new insights on:  1)~a large fraction of the
ultraluminous X-ray sources in nearby galaxies, 2)~gamma-ray bursts 
(GRBs) of long duration in distant galaxies, and 3)~the physics in the
jets of blazars. If jets in GRBs, microquasars and Active Galactic
Nuclei (AGN) are due to a unique universal  magnetohydrodynamic
mechanism, synergy of the research on these three different classes of 
cosmic objects will lead to further progress in black hole physics and
astrophysics.        

\keywords{Black hole physics -- X-rays -- binaries: jets -- stars: general}
\end{abstract}

\firstsection 

\section{Introduction}
The physics in systems that contain black holes is essentially the same, 
and  it is governed by the same scaling laws. 
The main differences derive 
from the fact that the scales of length and time of the phenomena 
are proportional to the mass of the black hole. If the lengths, 
masses, accretion rates, and luminosities are expressed in 
units such as the gravitational radius 
(R$_g$ = 2GM/c$^2$), the solar mass, 
and the Eddington luminosity, then the same physical laws apply to 
stellar-mass and supermassive black holes (Sams \etal\ 1998; Rees 2003). 
For a black hole of mass $M$ the density and mean temperature in the accretion 
flow scale 
with $M^{-1}$ and  $M^{-1/4}$, respectively. For a given
critical accretion rate, the bolometric luminosity and length of relativistic 
jets are proportional to the mass of the black hole. 
The maximum magnetic field at a given radius in a radiation dominated 
accretion disk scales with  $M^{-1/2}$, which implies that 
in the vicinity of stellar-mass black holes the magnetic fields may 
be 10$^4$ times stronger 
than in the vicinity of supermassive black holes (Sams \etal\ 1998). 
In this context, it was proposed (Mirabel \etal\ 1992; Mirabel \&
Rodr\'\i{}guez 1998) that 
supermassive black holes in quasars and stellar-mass black holes in 
X-ray binaries should exhibit analogous phenomena. Based on this 
physical analogy, the word 
``microquasar" (Mirabel \etal\ 1992) was chosen to designate compact 
X-ray binaries that are sources of relativistic jets (see Figure 1). 

\begin{figure}[tbh]
\includegraphics[width=0.6\textwidth]{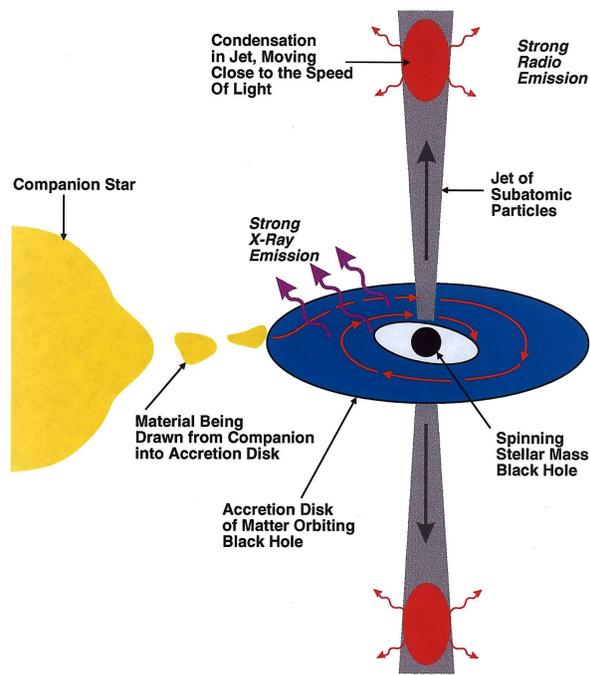}
\hfill
\parbox[b]{0.37\textwidth}{\caption{This diagram illustrates current ideas of 
what microquasars might be:  
compact objects -- black holes and neutron stars -- that are accreting mass from 
a donor star in X-ray binary systems and ejecting plasma at relativistic speeds.
The diagram not to scale.\newline}}
\end{figure}

\section{Superluminal motions in microquasars and quasars}
A galactic superluminal ejection was observed for the first 
time in the black hole X-ray binary GRS 1915+105,  
at the time of a sudden drop at 20--100 keV. 
Since then, relativistic jets with comparable bulk Lorentz factors 
$\Gamma$ = 1/[(1-$\beta$)$^2$)]$^{1/2}$ as 
in quasars have been observed in several other 
X-ray binaries (Mirabel \&
Rodr\'\i{}guez 1999; Fender 2002; Paredes 2004). At present, it is believed  
that all X-ray accreting black hole binaries are jet sources.

Galactic microquasar jets usually move in the plane of the sky 
$\sim$10$^3$ 
times faster than quasar jets and can be followed 
more easily than the later  (see Figure 2). 
Because of their proximity, in microquasars two-sided jets 
can be observed, which together with the distance provides 
the necessary data to solve the system of 
equations, gaining insight on the actual speed of the ejecta. 
On the other hand, in AGN located at 
$\leq$100 Mpc, the jets can be imaged with resolutions of a few times 
the gravitational radius of the supermassive black hole, as was done for 
M~87 (Biretta \etal\ 2002). This is  
presently not possible in microquasars, since such a precision in terms of 
the gravitational radius of a stellar-mass black hole would 
require resolutions a few hundreds of kilometers. Then, in terms of the 
gravitational radius
in AGN we may learn better  
how the jets are collimated close to the central engine. In summary, 
some aspects of the relativistic jet phenomena associated to accreting 
black holes are better observed in 
AGN, whereas others can be better studied in microquasars. 
Therefore, to gain 
insight into the physics of relativistic jets in the universe, synergy 
between knowledge of galactic and extragalactic black hole is needed.

\begin{figure*}[tbh]
\centering
\includegraphics[width=0.79\textwidth]{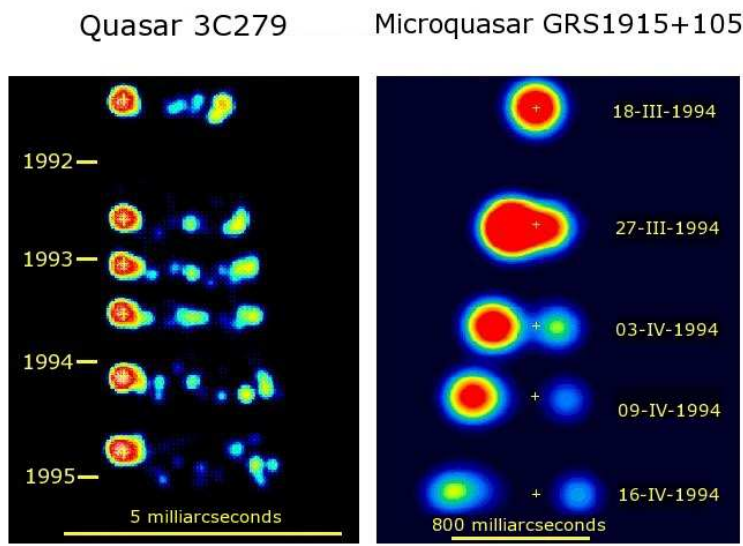}
\hfill
\parbox[b]{0.2\textwidth}{\caption{Apparent superluminal motions observed  
in the microquasar GRS~1915+105 at 8.6~GHz and in 
the quasar 3C~279 at 22~GHz.\newline}}
\label{mylabel2}
\end{figure*}

\section{Accretion-jet connection in microquasars and quasars}
Microquasars have allowed to gain insight into the connection 
between accretion disk instabilities and the formation of jets.  
In $\sim$1 hour of simultaneous multi-wavelength observations of GRS 1915+105 
during the frequently observed 30-40 min X-ray oscillations in this 
source, the connection between sudden drops of 
the X-ray flux from the accretion disk and the onset 
of jets were observed in several occasions (Mirabel \etal\ 1998; 
Eikenberry \etal\ 1998); see Figure~3. 

\begin{figure}[tbh]
\includegraphics[width=0.5\textwidth]{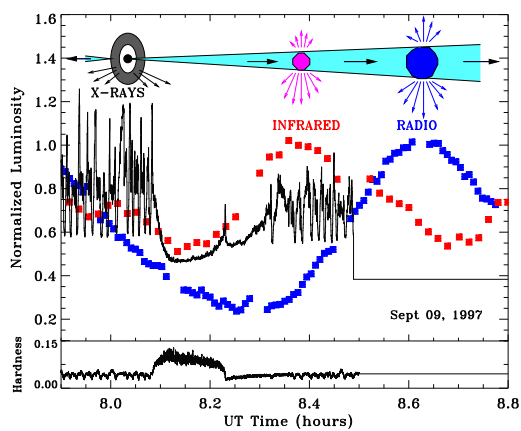}
\hfill
\parbox[b]{0.45\textwidth}{\caption{Direct evidence for the disk--jet connection 
in the microquasar GRS 1915+105 (Mirabel \etal\ 1998). When the hot inner 
accretion disk disappeared, its X-ray brightness abruptly diminished. 
The ensuing X-ray recovery documented the inner disk's replenishment, while 
the rising infrared and radio emission showed plasma being ejected in 
jet-forming episode. The sequence of events shows that material indeed 
was transfered from the disk to the jets. Similar transitions have been 
observed in the quasar 3C 120 (Marscher \etal\ 2002), but in time scales 
of years, rather than minutes.\newline}}
\end{figure}

From these observations we have learned the following:

\begin{description}
\item{}a) the jets appear after the drop of the X-ray flux;
\item{}b) the jets are produced during the replenishment of the inner 
accretion disk;
\item{}c) the jet injection is not instantaneous. It can last up to 
$\sim$10 min;
\item{}d) the time delay between the jet flares at wavelengths of 2$\mu$m, 
2cm, 3.6cm, 6cm, and 21cm are consistent with the model of 
adiabatically expanding clouds that had been proposed to account 
for relativistic jets in AGN (van der Laan 1966).
\item{}e) synchrotron emission is observed up to infrared wavelengths and 
probably up to X-rays. This would imply the presence of electrons in the jets with TeV energies;
\item{}f) VLBA images during this type of X-ray oscillations (Dhawan \etal\ 2000) showed that 
the ejecta consist of compact collimated jets with lengths of $\sim$100 AU;
\item{}g) there is a time delay of $\sim$5 
min between the large drop of the X-ray flux from the accretion 
disk and the onset of the jets. These $\sim$5 minutes of silence suggest 
that the compact object in GRS 1915+105 has a space-time border, rather than a material border, 
namely, a horizon as expected in 
relativistic black holes. However, the absence of evidence of 
a material surface in these observations could have alternative explanations.
\end{description}

After the observation of this accretion disk-jet connection in a microquasar, 
an analogous connection was 
observed in the quasar 3C 120 (Marscher \etal\ 2002), but in scales of 
years rather than minutes. This time scale ratio is 
comparable to the mass ratio between the supermassive black 
hole in 3C 120 and the stellar black hole in GRS 1915+105, as expected 
in the context of the black hole analogy.

\bigskip

\discuss{Phil Charles}{Could the apparent ``dark jet'' paradox of
SS433 be answered by the high inclination which causes obscuration of
most of the flux?}

\discuss{Felix Mirabel}{Yes, the accretion disc produces high opacity to
the X-rays and, as shown in the talk by Andrew King, the disc is
probably highly warped. The source could be very bright if seen from a
different direction.}

\discuss{Virginia Trimble}{Suppose there have never been those jets in
SS433, would there still be a supernova remnant there, or is the host
nebula blown mainly by the activity of central source?}

\discuss{Felix Mirabel}{This is an open
question. Clearly, the lateral extentions seen in the nebula W 50 that hosts
SS433, have been blown away by the jets.}

\discuss{Gregory Beskin}{Is it possible to create jets without a black
hole in the centre of the disc, for example by a neutron star?}

\discuss{Felix Mirabel}{Yes indeed, there are jets in binaries with
confirmed neutron stars. Clear cases are Scorpius X-1 and Circinus
X-1.}

\discuss{Gloria Dubner}{Comment on the previous question by V.~Trimble:
There is a way to probe if the whole bubble was created by a compact
object. I believe that the whole nebula had been created by a supernova
remnant and its shape was then distorted by the action of SS433's jets.
The way to address this issue is by searching for spectral changes in
the radio emission -- the jets have different spectrum than the rest of
the bubble.}
\cleardoublepage 

\vspace*{-1em}

\begin{thebibliography}{}

\bibitem[Biretta \etal\ 2002]{Biretta}
Biretta J.~A., Junor W., Livio M. 2002, NewAR, 46, 239

\bibitem[Dhawan \etal\ 2000]{Dhawan}
Dhawan V., Mirabel I.F., Rodr\'\i guez L.F. 2000, ApJ, 543, 373

\bibitem[Eikenberry \etal\ 1998]{Eikenberry} 
Eikenberry S.~S., Matthews K., Morgan E.~H., Remillard R.~A., Nelson R.~W. 1998, ApJ, 494, L61

\bibitem[Fender 2002]{Fender1}
Fender R. 2002, Lecture Notes in Physics, 589, 101

\bibitem[1998]{Marscher}
Marscher A.~P., Jorstad S.~G., G\'omez J.~L. \etal\ 2002, Nature 417, 625 

\bibitem[1998]{m98} 
Mirabel I.~F., Dhawan~V., Chaty~S. \etal\ 1998, A\&A, 330, L9

\bibitem[Mirabel \etal\ 1992]{Mirabelnat92}
Mirabel I.~F., Rodr{\'\i}guez L.~F., Cordier B. \etal\ 1992, Nature 358, 215

\bibitem[Mirabel \& Rodr{\'\i}guez 1998]{Mirabelnat98}
Mirabel I.~F., Rodr{\'\i}guez L.~F. 1998, Nature, 392, 673

\bibitem[Mirabel \& Rodr{\'\i}guez 1999]{MirabelARAA}
Mirabel I.~F., Rodr{\'\i}guez L.~F. 1999, ARAA, 37, 409

\bibitem[Paredes 2005]{Paredes}
Paredes J.~M. 2005, in Radio Astronomy from Karl Jansky to Microjansky,
eds. L.~I. Gurvits, S.~Frey \& S.~Rawlings (Budapest), EAS Publications Series, vol.~15, 
pp.~187--206 (astro-ph/0402671)

\bibitem[van der Laan 1966]{vanderLaan}
van der Laan H. 1966, Nature, 211, 1131

\bibitem[Rees 2003]{Rees}   
Rees M.~J. 2003, in Future of Theoretical Physics and Cosmology, ed. G.~W. Gibbons \etal\
(Cambridge University Press: Cambridge), pp.~217--235  (astro-ph/0401365)

\bibitem[Sams \etal\ 1998]{Sams}
Sams B.~J., Eckart A., Sunyaev R. 1998, Nature, 392, 673

\end{thebibliography}
\end{document}